\documentclass[twocolumn]{aa}
\usepackage{graphicx}
\usepackage{txfonts}
\usepackage{supertabular}
\usepackage{natbib}
\usepackage{isolatin1}
\usepackage{rotate}
\input{journals.defs}

\begin{document}
   \title{Automatic classification of eclipsing binaries light curves using neural networks}
   \titlerunning{Automatic classification of light curves.}

   \author{Sarro$^{1}$, L.M.
          \and
          Sánchez-Fernández$^{2,3}$, C.
          \and
          Giménez$^{4}$, Á.
        }

   \offprints{L.M. Sarro}

   \institute{Dpt. de Inteligencia Artificial, U.N.E.D., c/ Juan del Rosal, 16, 28040 Madrid, Spain\\
              \email{lsb@dia.uned.es} Telephone: 00 34 91 3988715 Fax: 00 34 91 3988895
         \and
         Laboratorio de Astrof\'{\i}sica Espacial y F\'{\i}sica Fundamental, P.O. Box 50727, E-28080 Madrid, Spain
         \and
         XMM-Newton SOC, ESAC, P.O. Box 50727 E-28080 Madrid, Spain\\
             \email{Celia.Sanchez@sciops.esa.int}
         \and
         Research and Scientific Support Department, ESA, ESTEC, Postbus 299, 2200 AG Noordwijk, 
         The Netherlands\\ \email{agimenez@rssd.esa.int}
          }

   \date{}

   \abstract{
     
     In this work we present a system for the automatic classification
     of the light curves of eclipsing binaries. This system is based on a
     classification scheme that aims to separate eclipsing binary
     sistems according to their geometrical configuration in a
     modified version of the traditional classification scheme. The
     classification is performed by a Bayesian ensemble of neural
     networks trained with {\em Hipparcos} data of seven different
     categories including eccentric binary systems and two types of
     pulsating light curve morphologies. 
            }

   \maketitle
%
%________________________________________________________________

\section{\label{1}Introduction}

Eclipsing binaries (hereafter EBs) play a fundamental role in modern
astrophysics for several reasons. First of all, detached double-lined
EBs without mass transfer between the components are a prime tool to
derive fundamental stellar parameters; joint analysis of their
light and radial velocity curves provides accurate (1--2\%)
determinations of masses, radii, and luminosity ratios.  Eclipsing
binaries also work as testing grounds for stellar structure and
evolution models, and as such they play a key astrophysical role
across the whole HR diagram.  Recently, the study of EBs in other
galaxies and clusters has made it possible to explore stellar
evolution and to establish mass-luminosity laws for galaxies with a
vastly different evolutionary and chemical histories from our own
Galaxy (such as LMC and SMC). Moreover, EBs are beginning to play an
important role in cosmology as distance indicators to nearby galaxies.
Studies of Galactic early-type binaries have shown that distance
moduli accurate to $\pm$ 0.1 mag are attainable, a precision
comparable to that obtained for individual Cepheid variables. As more
data are accumulated, studies of these systems may lead to an
improvement in the extragalactic distance scale.

In recent years, large scale photometric surveys have been providing a
wealth of light curves of variable stars out of which a large amount
of EB systems can be selected.  For example, the ESA astrometric
satellite {\em Hipparcos} found 70\% new variables out of the
relatively bright selected sample.  The {\em GAIA} large-scale
photometric survey will also have significant scientific value for the
study of nearly all types of variable stars, including eclipsing
binaries. It is expected that about 1 million EBs, those with $V\leq16$
mag, will be discovered.  Even if reliable physical parameters could
be derived for only 1\% of the observed EBs, this would be a great
contribution to stellar astrophysics in comparison with what has been
obtained so far from ground-based observations. The Optical Monitoring
Camera (OMC; \citeauthor{2003A&A...411L.261M},
\citeyear{2003A&A...411L.261M}) onboard {\em INTEGRAL} is another
example of an instrument that continuously provides high quality
photometric measurements of thousands of eclipsing and pulsating
variables, amongst other objects more closely related to high energy
astrophysics. Finally, the COnvection ROtation and planetary Transits
({\em COROT}) mission will produce, as a by product, enormous amounts
of light curves of objects with unprecedented accuracy (see e.g.
\citeauthor{2002sshp.conf...17B} \citeyear{2002sshp.conf...17B}). All
these vast amounts of data offer the opportunity to select not only
EB light curves, but all kinds of light curves for deeper
investigation and/or follow-up. Such databases also provide
astronomers with powerful heuristics like the possibility to construct
statistically significant samples of objects that can be used as
probes for correlations between physical parameters, e.g., in the case
of the rotation-activity correlation \citep{MontesinosandJordan}.
Nevertheless and despite all this encouraging prospects, it is
becoming increasingly clear that intelligent processing of these large
datasets is needed, and no method based on manual procedures can be
used. It is precisely the enormity of this volume of data that makes
it necessary to implement automatic light curve classification tools
before any serious scientific analysis. Fortunately, it is exactly
in these kinds of tasks (such as pattern recognition, classification,
clustering, and knowledge discovery in the form of dimensional
correlations) that machine learning and artificial intelligence
techniques yield their best performances. 

In this paper we concentrate on the applications of neural networks
for the task of light curve identification and clustering. Neural
networks have been widely used in the past for classification of
stellar \citep{2001ApJ...562..528S} and galactic
\citep{1996MNRAS.283..651F} spectra, star/galaxy separation in images
(\citeauthor{2002A&A...385.1119P}, \citeyear{2002A&A...385.1119P};
\citeauthor{2001ApJ...556..937C}, \citeyear{2001ApJ...556..937C}), or
quasar detection. Sometimes the classification process takes input
data spanning a combination of spatial and temporal dimensions as in
the case of solar flare detection where time series of images are used
for the classification of events \citep{2002SoPh..206..347F}. They
have also been used for time series prediction
\citep{2000SoPh..191..419V}, nonlinear system identification
(\citeauthor{2000A&A...357..197B}, \citeyear{2000A&A...357..197B};
\citeauthor{2001A&A...378..316C}, \citeyear{2001A&A...378..316C}), and
telescope control \citep{1991Natur.351..300S} to cite but a few. In
the specific field of light curve analysis, neural networks have been
used recently for clustering purposes \citep{2004MNRAS.353..369B} and
for microlensing detection (\citeauthor{2003MNRAS.341.1373B}
,\citeyear{2003MNRAS.341.1373B}, \citeauthor{2004MNRAS.352..233B}
,\citeyear{2004MNRAS.352..233B}). Here we present a refined classifier
for eclipsing binaries based on state-of-the-art neural networks that
builds upon some of the work presented in these previous developments.

In this work we apply Bayesian techniques to the training of neural
networks for the automatic classification of light curves of variable
stars based solely on their morphological aspects. The network is able
to recognize four types of eclipsing binary systems and two types of
pulsating star light curves. Furthermore, all the types define a link
between the morphology of the light curve and the underlying physical
scenarios as much as possible. In Sect. \ref{2} we describe the
classification scheme in detail; in Sect. \ref{3} we describe the
preprocessing of the data and the neural network architecture and
training; in Sect. \ref{4} we describe the results obtained, assess the
quality and performance of the system and analyse the resulting
connection topology; finally, in Sect. \ref{5}, we summarise the
conclusions of this work.

%__________________________________________________________________

\section{\label{2} Classification scheme}

One obvious requirement of any classification system is the direct
link between the features used as input and the classes defined from
them. In the realm of variable systems, unfortunately, we find that
either the classes established up to now are not consistently defined
in terms of the light curves, as in the case of eclipsing binaries, or
there are degeneracies, as in the case of pulsating stars, in the sense
that different categories can have morphologically identical light
curves. With pulsating variable light curves, the degeneracy can only
be resolved with supplementary spectral information and periods. This
problem will be addressed in a future paper where a multi-agent expert
system will be presented, which is capable of classifying pulsating stars
(identified by their light curves) by navigating the Virtual
Observatory space searching for discriminant observations. Here we
restrict ourselves to the problem of separating pulsating variables
from eclipsing binary systems and subclassifying the latter into
physically inspired classes univocally defined in terms of their light
curves.

The two main factors that determine the shape of the light curve of an
eclipsing binary system are its geometric configuration (i.e. the size
of the component stars relative to their Roche lobes), which determines
the fraction of the light curve occupied by eclipses, and the relative
brightness of the stellar components, which determines the eclipse
depths.  The inclination of the system with respect to the line of
sight can affect the depths of the eclipses, but its effect on the
overall light curve morphology is less important. 

We propose here a classification scheme which aims to separate
eclipsing binary systems according to their geometrical
configuration. This scheme is adapted from the historical
classification of eclipsing binary light curves into three groups
(Algol, Beta Lyrae, and W Ursae Majoris), but attempts to solve the
problems of class heterogeneity and subjectivity of the traditional
light curve classification, which includes systems with different
physical properties in the same group. Our classification relates the
groups established to the geometry, in the sense that systems with
the same geometrical configuration are classified in the same group.

We note here a previous attempt to solve the degeneracy of the
traditional classification of eclipsing binaries light curves by
\cite{1997AJ....114..326A}. They proposed a decimal classification
scheme based on combining the relative radii of the stars and the
surface relative flux ratio. As an alternative, when only the light
curve morphology is available, we found that a simple 4-group
classification scheme suffices to separate the systems into
homogeneous classes.

\subsection{Our classification}

Definition of the classes %in terms of the morphological criteria
assumes that the light curve has been processed such that the phase of
the deeper eclipse (we refer to primary eclipse) is defined as
$0.75$.  Systems are classified into 4 groups as follows:

\begin{itemize}
\item{Class 1 systems: light curves with well-defined start and end to
both eclipses. These light curves may present small curvatures out of
eclipses, but this curvature never masks the beginning and end of the
occultations.} 
\item{Class 2 systems: light curves with only a well-defined primary
eclipse, while the secondary has no clear beginning or end.}
\item{Class 3 systems: light curves with eclipses of different depth
and no flat light curves out of ecplise. In these systems, the light
curve curvature out of eclipse masks the beginning and end of the
occultations.}
\item{Class 4 systems: light curves with the equal depth eclipses
alternating, and no flat light curve out of eclipse.}
\end{itemize}

\begin{figure*}[h]
%\centering
\includegraphics[scale=0.35,angle=-90]{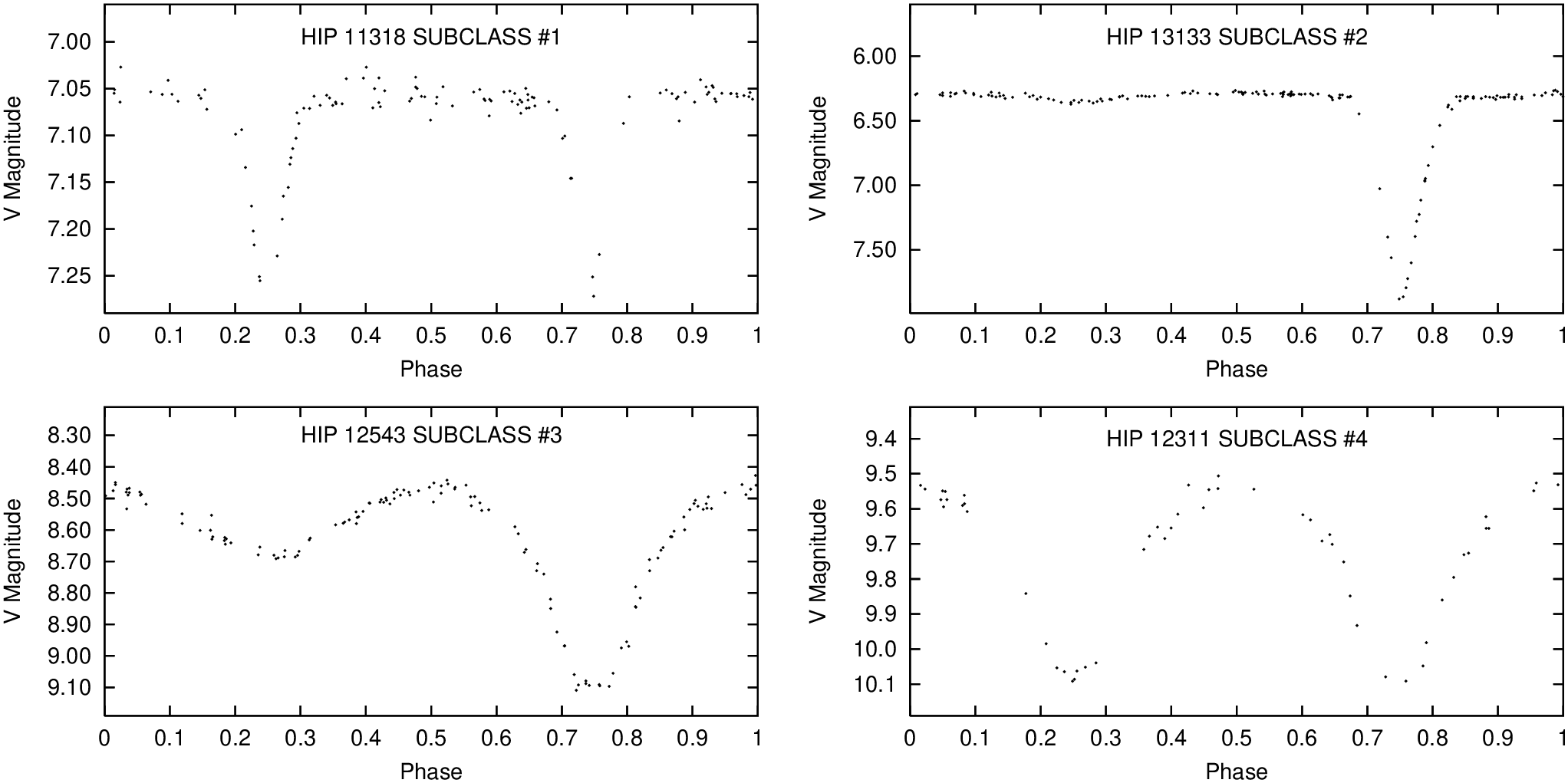}
\caption{\label{examples} Examples of the classes defined in the text obtained by Hipparcos, folded with the periods provided in the mission catalogue.}
\end{figure*}

Figure \ref{examples} shows example light curves from the Hipparcos
catalogue for each class.

% Geometric & evolutionary implications

\subsection{Application to a sample of systems}

In order to show the relation between the classes established by this
scheme and some of the system parameters, we classified a set of 81
binary systems with well-studied light curves and precisely determined
physical parameters. The list of systems used in this study and their
main physical parameters can be found in Tables
\ref{Tipo1}--\ref{Tipo4}.

\begin{figure}
\begin{center}
\includegraphics[scale=0.7]{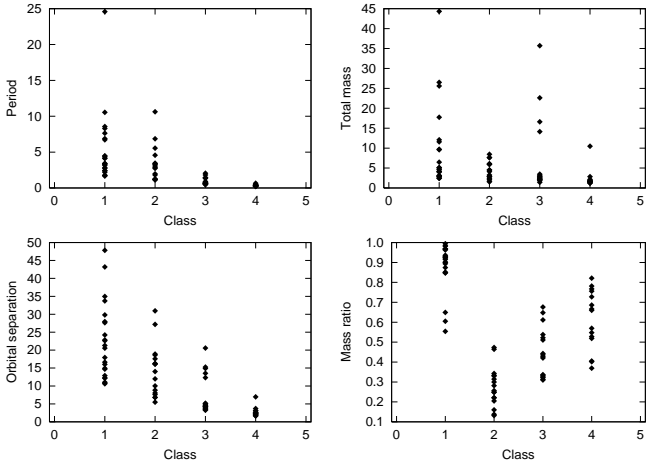}
\end{center}
\caption{System parameters grouped by class as defined in the
  text. Orbital periods (days) and separations (expressed in solar
  radii) are shown in the left column of the plot; total mass of the
  system (expressed in solar masses) and mass ratios are shown in the
  right column plots.}
\label{phys-by-group}
\end{figure}

In the following, we will analyse the classes in terms of the
component masses, orbital separation, mass ratio, and filling factors
of the 81 systems included in our sample. In order to help with the
interpretation of the combination of any two such parameters, we first
show in Fig. \ref{phys-by-group} the sample masses, orbital
separations and, mass ratio for the systems classified in each
class. We can see in the total mass plot that there is no discriminant
boundary or general trend between classes, although type 4 systems
seem to be characterized by a lower mass. Orbital separations, on the
other hand, show a decreasing trend towards higher types. We see how
an apparent segregation in the total masses of type 3 systems into two
sets (low and high mass objects) is reproduced in the separation plot
in the sense that the less massive systems also have lower orbital
separations and {\em vice versa}. This segregation into two groups may
be an artifact caused by a limited sample size. Thus, more systems of
this class with accurately determined parameters are needed to clarify
whether two different populations with similar light curves indeed
exist or whether there is continuous transition. Finally, the mean
value of the mass ratio of the components shows values closer to 1 for
type 1 systems, lowest values for type 2, and increasing values
thereafter (types 3 and 4).

\begin{figure}
\begin{center}
\includegraphics[scale=0.7]{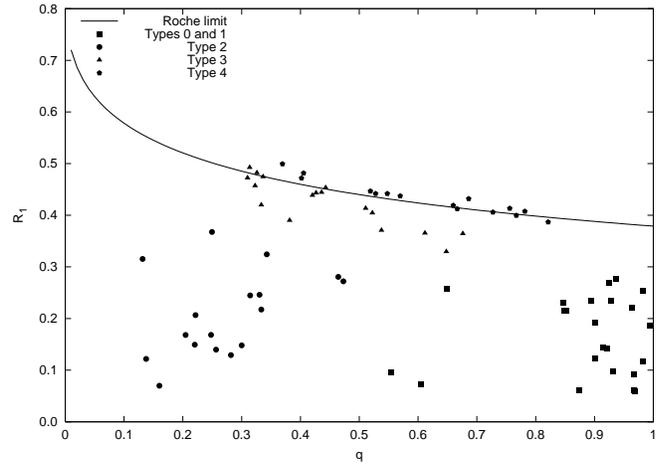}
\end{center}
\caption{Radius to orbital separation ratio of the primary components 
  in the sample as a function of the mass ratio $q$}
\label{radii-massratio1}
\end{figure}

\begin{figure}
\begin{center}
  \includegraphics[scale=0.7]{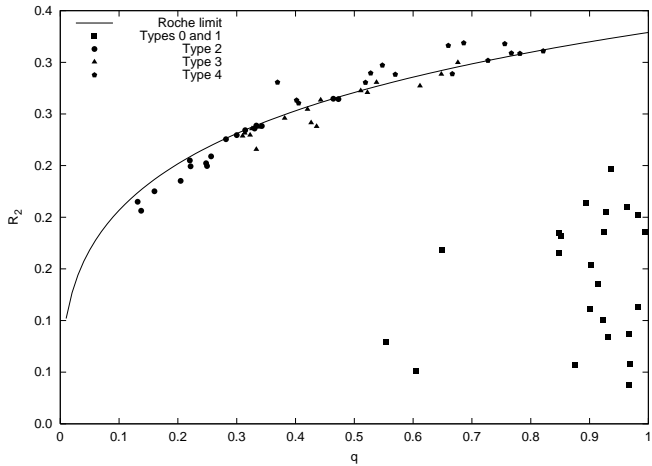}
\end{center}
\caption{Radius to orbital separation ratio of the secondary
components in the sample as a function of the mass ratio $q$}
\label{radii-massratio2}
\end{figure}

If we now plot the radius to orbital separation ratio for both
components of each system as a function of the mass ratio $q$ (defined
as $q=M_2/M_1$, $M_1$ being the most massive star), we obtain Fig.
\ref{radii-massratio1} and \ref{radii-massratio2} where we have also
included the Roche lobe size (in orbital separation units) computed
using the approximation of \cite{Eggl1983}. Figure
\ref{radii-massratio1} clearly shows that primary components of type 1
and 2 systems are well below the Roche lobe radius, while type 3
primaries are close to it, and type 4 primary stars clearly fill their
Roche lobes. At the same time, only radii of type 1 system secondary
stars are clearly below the Roche limit. Under this perspective, it is
evident that our classification scheme is a morphological
transposition of the different geometrical configurations: type 1
systems are composed of two stars with radii clearly below the Roche
lobe (detached systems); type 2 systems are composed of a primary star
well below its Roche limit and a secondary filling its Roche lobe
(i.e. semidetached systems); type 3 systems have a primary component
close to filling its Roche lobe and a secondary component already
filling the critical lobe and therefore, they represent semidetached
systems close to contact; finally, type 4 light curves represent
contact binaries with both components filling their Roche lobes and
possibly exceeding them. We will pursue further the implications of
this scheme after considering possible correlations between total
mass, orbital separation and, mass ratios.

\begin{figure}
\begin{center}
  \includegraphics[scale=0.7]{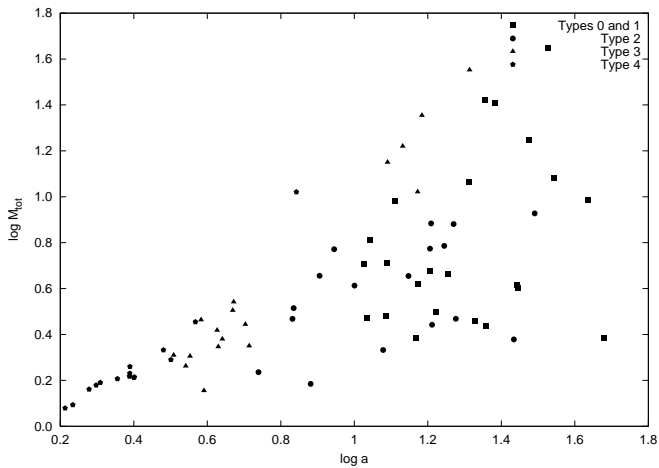}
\end{center}
\caption{Logarithmic representation of the eclipsing binary system total mass ($M_{tot}$) as a funtion of the orbital separation $a$ in solar radius units.}
\label{a-M}
\end{figure}

Figure \ref{a-M} represents all systems in the sample in the $\log
(M_{tot})$--$\log (a)$ space, with $M_{tot}$ the system total mass and
$a$ the orbital separation in solar radius units. Although there is
clearly no separability in this space, there are evident trends in the
data. Again, type 4 systems are found in the low orbital separation
and low total mass region of the plot, and seem to follow a tight
linear relation. The rest of the types continue this correlation with
increasing values of the dispersion: low mass type 3 systems follow
the trend to the right with higher values of both parameters, then
type 2 systems, and finally, with a high degree of overlapping, type 1
systems occupy the high total mass, high orbital separation region of
the plot.

Once revised, the physical characteristics of the proposed classes, we
can reformulate the definitions, this time summarizing the regions of
parameter space where we can expect to find the system.

\subsubsection{Systems with type 1 light curves}

These are detached systems with widely varying total masses and a wide
range of spectral types from O to F. Most of the systems assigned to
this class have mass ratios close to 1 due to selection effects. Both
components are well within the Roche lobe and have orbital separations
in the 10--100 $R_{\odot}$ range. All these properties result in light
curves with well-defined beginnings and ends of both eclipses and flat
regions outside them. These binaries are the best source of
information to study stellar absolute dimensions and structure.  Most
systems included in this group are eccentric.
 
\subsubsection{Systems with type 2 light curves}

Systems classified as type 2 have low mass ratios and total masses
below $10 M_{\odot}$. In the systems studied, the primary component of
spectral type A or B is in the Hydrogen burning phase, and the
secondary component (that originally transferred a significant
fraction of its mass through the $L_1$ Lagrange point ) has spectral
type G or K and has a mass around or below the solar mass. These
systems have their origin in detached systems in which the most
massive star evolved out of the main sequence, filled its Roche lobe,
and transferred mass to the secondary component through $L_1$ until
the original secondary became the most massive component. Although the
details of the process remain unclear (\citeauthor{Hall1975},
\citeyear{Hall1975}; \citeauthor{Ziol1976}, \citeyear{Ziol1976}),
system mass loss cannot be discarded.

\subsubsection{Systems with type 3 light curves}

Type 3 systems in the sample show two different clusters in parameter
space. They are all semidetached systems with the primary component
close to filling the Roche lobe and, due to the proximity of the
components, both eclipses alternate without intereclipse flat
intervals.

The most populated cluster is composed of systems with total masses
lower than 5 $M_{\odot}$ and is characterized by short period (less
than 1 day) orbits and small orbital separations in close-to-contact
configurations. The primary components are spectral types A or F, and
the secondary stars are one or two types cooler. 

The second, less populated cluster of systems corresponds to total
masses above 10 $M_{\odot}$ and periods in the 1-3 days range. They
show moderate mass ratios, and both components are of similar spectral
types around B. Again, the more massive primary component is close to
its Roche lobe but separated from it, and the evolved secondary is in
contact with its lobe. They probably originated in very close orbits,
with mass ratios around 1, and evolved to near contact configurations
as the stars expanded as a consequence of Main Sequence evolution.

\subsubsection{Systems with type 4 light curves}

All light curves classified as type 4 correspond to systems in contact
where both stars fill and possibly exceed their Roche lobes. If they
exceed this limit, the size of the common envelope depends on the most
external contact surface. These systems are characterized by short
periods, small orbital separations, and a wide range of mass ratios.
Except for RZ Pyx with a spectral type B, the rest of the systems are
composed of late type stars with total masses below 3 $M_{\odot}$. It
is not yet clear whether these systems are formed as contact binaries
or if they evolve from detached systems through loss of angular
momentum. Most possibly, the population of contact binaries is a
mixture of both evolutionary paths.

To finish this section we would like to point out that, unfortunately,
our scheme is not without degeneracies or cross-class contamination.
The main sources of contamination arise from pre-main sequence
detached systems. In these systems, one of the components is in the
contracting phase towards the Main Sequence while the second component
has already stabilized in it. The former is far dimmer than the latter
and can thus be a source of confusion with type 2 systems despite
their detached geometry. Due to the relative youth of these pre-main
sequence systems they generally have not had enough time to
circularise their orbits and show therefore some degree of
eccentricity. This criteria can be used to place them correctly in the
type 1 group. Nevertheless, certain orientations of the orbit with
respect to the observer may result in eclipses being in quadrature
despite the eccentricity of the system.

\section{\label{4}Results and discussion.}

In order to assess the performance of the ensemble of neural networks
thus generated, we divided the whole set of examples into two groups:
(i) a training set used to obtain the {\em a posteriori} probabilities
of the parameter sets generated by MCMC methods (75\% of the complete
set), and (ii) a test set used to obtain estimates of the expected
cross-class misclassification rates (25\% of the complete set). In
order to approximately maintain the relative size of each class in the
complete set, a light curve is assigned to the training set with a
0.75 probability or to the test set with a 0.25 probability. This
splitting is performed 10 times and the resulting blocks considered
separately. Errors in the performance estimates correspond to the
root sum square of the performance of the ten
partitions. Furthermore, three different network architectures are
tested.  Invariably, all three architectures have a 50-unit input
layer and a 7-unit output layer.  They differ on the presence/absence
of one or several hidden layers.  The first network is a logistic
regression network (with no hidden layer); the second network has one
hidden layer with 30 units; and the third network architecture
contains two hidden layers of 20 and 10 units, respectively. Each
splitting of the complete set is used to generate 1000 networks of
each architecture, and the last 200 are used to predict classes for
the light curves in the corresponding test sets. Thus, we end up with
10$\times$3 ensembles of 200 neural networks. In addition to this, the
effectiveness of the ARD procedure was assesed by comparing the
predictions on the test sets of networks of the same architecture
with/without ARD implemented in the training process.

In order to avoid unnecessary computations, we checked the average
error percentage for each architecture and found an $8.7\%\pm3.6$ for
the 50-7 architecture, $6.9\%\pm1.3$ for 50-30-7 and $6.9\%\pm1.3$ for
50-20-10-7. The average log probability of the test cases was -0.24
for the 50-7 network, -0.17 for the 50-30-7 network, and -0.18 for the
50-20-10-7 network.  Although ARD could naturally prune unnecessary
units and connections, if hyperparameters were introduced in all
network layers, we preferred to continue the analysis with the 50-30-7
architecture.

The performance of the neural network does not depend on the
total number of measurements in the light curve. It would indeed
depend on the total information content of the available points (note
that the total information content combines information not only on
the phase coverage but also on the relevance of the covered phases for
classification purposes), if no pattern completion were carried out
during the preprocessing stage. This can be seen by taking the extreme
case of an infinite number of points concentrated on a very narrow
phase interval where all classes present the same behaviour. But, as
explained above, the preprocessing stage completes the missing bins
using the curvature of the closest light curves in the SOM. Therefore,
if the completion process is correct and the initial incomplete light
curve has enough information to reconstruct the missing phases, no
dependence of the neural networks performance on the information
content of the light curve before preprocessing should be detected,
which is in fact the case down to the minimum information content
found on Hipparcos light curves; around 20\%, 100\% is a complete
light curve. Unfortunately, this robustness is not realistic since
only a 10\% of the catalogue has information content below 60\%, and
therefore the statistics are rather poor. The study was carried out
grouping the light curves in bins of information content width 20\%.
The smaller number of cases in each bin increases the standard
deviation up to 4.3\%.

We also investigated the dependence of the classifier performance
on the ratio between the amplitude and the errors in the measurements
of the light curve (the signal-to-noise ratio) and found no
significant trend above a mean variance of 5.7\%. Again, the
preprocessing stage tries to minimize the effect of the errors in the
measurement by means of the regression process. It has to be beared in
mind that it is actually a smooth curve (the result of the second
regression) that is used as input to the neural net.

The robust performance of the neural network described in the
preceding paragraphs can also be expected for light curves in the
same information content and signal-to-noise ratio ranges as those
found in the Hipparcos catalogue. As mentioned above, this implies
light curves with information contents above 60\% (although the
classifier shows the same performance down to a 20\% with only a few
tens of light curves to compute the means). Light curves with lower
information contents can possibly be mistankenly completed if not
enough information is available for a reliable completion. Regarding
the signal-to-noise ratios, we have found that 98.5\% of the light
curves in the catalogue have ratios above 5$\sigma$.

Finally, we studied the performance of the neural network as a
function of the number of bins used as input and found that 50 bins
lies in a plateau with similar performances that goes from 40 bins up
to 90 bins. Below 40 bins the degradation is first due to the
misclassification of eccentric systems and below 20 bins mainly to
confusion between types 1 and 3. Above 90 bins, there are not enough
examples to construct the relationship between each input node and the
class and the performance curve begins a slow decline as expected.

Table \ref{misclass-percent} shows the average cross-class
misclassification percentage and the standard deviation computed for
the 10 different splittings of the complete set for the 50-30-10
architecture. Each row lists the percentage of objects of a given type
that have been misclassified in all other possible categories.

% Old data w/o synthetic LCs Simple average
%\begin{table}[]
%\caption[]{.}
%\label{misclass-percent}
%\centering
%\begin{tabular}{c c c c c c c c}
%\hline
%& Type 0 & Type 1 & Type 2 & Type 3 & Type 4 & Type A & Type B\\
%\hline
%\hline
%Type 0  & -   & 12.4 $\pm$ 3.6 & 8.5$\pm$4.9  & 1.1$\pm$1.1   & 0.0$\pm$0.0  & 0.0$\pm$0.0   & 0.0$\pm$0.0     \\
%Type 1  & 0.5$\pm$0.2  & -     & 1.4$\pm$0.6  & 5.4$\pm$0.9   & 1.7$\pm$0.5  & 0.0$\pm$0.0   & 0.0$\pm$0.0     \\
%Type 2  & 0.0$\pm$0.0  & 3.0$\pm$0.9   & -    & 0.0$\pm$0.0   & 0.0$\pm$0.0  & 0.0$\pm$0.0   & 0.0$\pm$0.0     \\
%Type 3  & 0.0$\pm$0.0  & 5.4$\pm$1.1   & 0.0$\pm$0.0   & -    & 3.6$\pm$0.8  & 0.0$\pm$0.0   & 0.6$\pm$0.5     \\
%Type 4  & 0.0$\pm$0.0  & 2.1$\pm$0.5   & 0.0$\pm$0.0   & 5.1$\pm$1.6   & -   & 0.0$\pm$0.0   & 0.0$\pm$0.0     \\
%Type A  & 0.0$\pm$0.0  & 0.0$\pm$0.0   & 0.0$\pm$0.0   & 0.0$\pm$0.0   & 0.0$\pm$0.0  & -    & 9.8$\pm$1.8     \\
%Type B  & 0.0$\pm$0.0  & 0.1$\pm$0.1   & 0.0$\pm$0.0   & 0.0$\pm$0.0   & 0.0$\pm$0.0  & 3.4$\pm$0.4  & -       \\
%\hline
%\end{tabular}
%\end{table}
%

% Data with Hipp light curves + synthetic eccentric LCs Weighted average
\begin{table*}[]
\caption[]{Cross-class misclassification percentages. Each row lists the percentage of light curves of a given class that were mistakenly classified as belonging to the corresponding type in the row of headers.}
\label{misclass-percent}
\centering
\begin{tabular}{c c c c c c c c}
%\hline
& Type 0 & Type 1 & Type 2 & Type 3 & Type 4 & Type A & Type B\\
\hline
\hline
Type 0  & -   & 3.6 $\pm$ 0.6 & 0.5$\pm$0.3  & 0.3$\pm$0.3   & 0.0$\pm$0.0  & 0.0$\pm$0.0    & 0.0$\pm$0.0     \\
Type 1  & 1.0$\pm$0.3  & -    & 1.5$\pm$0.4  & 4.9$\pm$0.7   & 1.6$\pm$0.3  & 0.0$\pm$0.0    & 0.0$\pm$0.0     \\
Type 2  & 0.2$\pm$0.2  & 4.0$\pm$0.9   & -   & 0.0$\pm$0.0   & 0.0$\pm$0.0  & 0.0$\pm$0.0    & 0.0$\pm$0.0     \\
Type 3  & 0.0$\pm$0.0  & 8.8$\pm$1.7   & 0.0$\pm$0.0   & -   & 5.3$\pm$1.1  & 0.0$\pm$0.0    & 0.0$\pm$0.0     \\
Type 4  & 0.0$\pm$0.0  & 2.1$\pm$0.7   & 0.0$\pm$0.0   & 7.4$\pm$1.0   & -  & 0.0$\pm$0.0    & 0.0$\pm$0.0     \\
Type A  & 0.0$\pm$0.0  & 0.0$\pm$0.0   & 0.0$\pm$0.0   & 0.0$\pm$0.0   & 0.0$\pm$0.0  & -    & 13.4$\pm$1.7     \\
Type B  & 0.0$\pm$0.0  & 0.2$\pm$0.1   & 0.0$\pm$0.0   & 0.0$\pm$0.0   & 0.0$\pm$0.0  & 3.2$\pm$0.4  & -       \\
\hline
\end{tabular}
\end{table*}

%ARD discussion

\begin{figure}[]
\begin{center}
\includegraphics[scale=0.7]{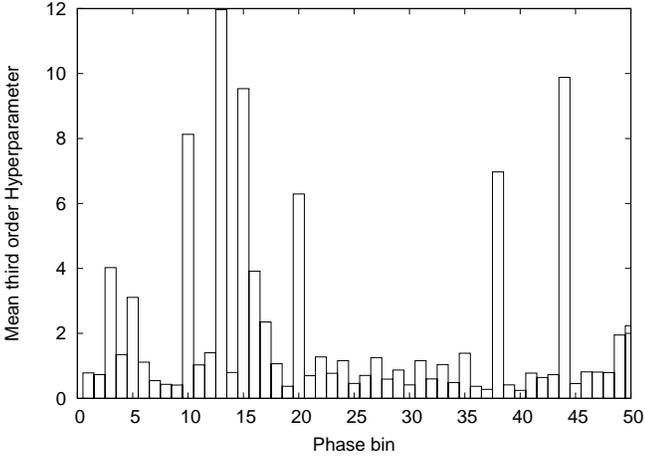}
\end{center}
\caption{\label{ARD} Mean values of the hyperparameter controlling the average magnitude of the 
  weights out of the input unit for the Hipparcos set plus 112
  synthetic light curves of eccentric systems.}
\end{figure}

These percentages are less than 1 point lower on average than obtained
without ARD implemented during the training. From this point onward,
and having an estimate of the expected misclassification rate, we
continue the analysis of the performance of our classification system
with the complete set of Hipparcos plus synthetic eccentric light
curves as training set of a 50-30-7 architecture network with ARD
implemented. Although this particular choice is only marginally
justified in terms of classification performance, we consider that it
provides best results with maximum information. Figure \ref{ARD} shows
the mean values of the third level hyperparameter controlling the
average strength of the connections out of each input unit, obtained
in the last 1500 networks with ARD implemented.

It shows non negligible values at all phase bins although the average
strength of neural connections from input units is seen to display
some degree of structure. On a background of low magnitude weights, we
find a higher concentration of sensitive units around phase 0.25. This
can be easily understood in terms of the classification criteria
exposed in Sect. \ref{2} and the preprocessing of the light curves,
the combination of which makes the class assignment decision depend
mainly on the properties of that region. The relative importance of
the connection strength of synapsis out of the unit representing phase
$\phi=0.75$ can be explained under the ARD framework as the result of
the MCMC methods blindly exploring the hyperparameter space with a
probability given only by the prior.  This unit conveys no information
at all (the preprocessing stage fixes its input at 1.0) and therefore,
we can expect its posterior probability to be roughly equal for all
possible values of this hyperparameter. The final value is simply an
average of this blind exploration of the prior. The low sensitivity of
input units away from $\phi=0.25$ can be explained by the easy
separability of very eccentric systems with respect to all other
classes.

\section{\label{5} Conclusions}
In this work we present an automatic light curve classifier based on
neural networks able to separate pulsating stars from eclipsing binary
systems. We classified the latter into 4 groups, according to a new
classification scheme based solely on the morphological features of
the light curve, which maps the system geometrical configuration.  We
applied the new classification scheme to a sample of 81 systems with
well-measured light curves and well-determined physical parameters, to
investigate the physical properties of the classes thus defined. We
found that, based only on the light curve morphology, we are able to
separate systems with different geometrical configurations.

From a technical point of view, the improvement of our classification
scheme with respect to the traditional one relies mainly establishing
well-defined and objective criteria that can be easily implemented on
a neural network.  The traditional classification was not
systematically formulated and was subjectively applied after visual
inspection of the light curve by the observer. From a physical point
of view, our classifcation scheme improves the traditional
classification by establishing classes characterized by the variation
of the system geometry from one group to the other. In the traditional
one, systems with different geometrical configurations were classified
in the same group.

We also considered under what circumstances the classification
scheme proposed here would fail {\em a priori} to reflect the
underlying geometrical configuration and found the following two
exceptions:

\begin{itemize}
\item{Pre-main-sequence systems with low luminosity secondaries and
  semimajor axis aligned with the visual line.}
\item{Semi-detached systems with a secondary component in contact with
  the Roche lobe and a primary close to contact, being both stars of
  the same or similar luminosities.}
\end{itemize}

We explored the classifier performance when trained with Hipparcos
examples alone and together with a set of light curves artificially
generated to increase the relative frequency of eccentric systems in
the training set. In the latter case we found a significant
improvement in the classifier's ability to detect eccentric binary
systems at the expense of a small degradation in the overall
performance. We explored several architectures for the network and
found improved performance for networks with one or more hidden layers
(with negligible differences between them). Finally, we also found
negligible differences between the performance of classifiers trained
with and without ARD.  Almost the entire set of misclassifications
occur at the boundaries between classes mainly due to the
nonseparability of the sets of examples. We atribute this to the
presence of noise in the training set.  Nevertheless the softmax
formulation of the model provides a quantitative measurement of the
confidence in the class assignment in such cases, because systems in
the proximity of a boundary between two classes exhibit comparable
values in the output of the neurons that label those classes.

We have compared the performance of Bayesian neural networks
presented above with that of a simple multilayer perceptron and found
an overall improvement of 12.1\%; i.e., the percentage of right
classifications of a 50-30-7 multilayer perceptron is 19.0\%. These
figures combine both the improvement in the regression stage and that
in the final classification. The inclusion of wide priors for the
hyperparameters leads to increased robustness when outliers are
expected. In our case, we have experienced that the results of
regression with simple multilayer perceptrons, in the presence of
outliers to the light curve, are significantly worse than those of
Bayesian neural networks.

\bibliographystyle{aa.bst} 
\bibliography{eclann}

\Online

\section{\label {3} Neural classifier}

As mentioned in the introduction, the final aim of this work is to
make the computational classification of automatically preprocessed
light curves possible without human supervision. The classification
system defined in the previous section was designed to accomplish this
goal, while at the same time preserving the physical significance
investigated there. In this section, we describe the methodology used
to implement the classifier and the results obtained as assesed using
standard techniques in the field of connectionism.

\subsection{Bayesian training of neural networks}

Most connectionist methods consist of distributing the computation of
the solution of a given task amongst a number of interconnected,
formally equivalent units or neurons performing very simple nonlinear
operations upon the weighted sum of their inputs. The connection
topology divides the ensemble of neurons (the neural network) in
layers with forward connectivity. This architecture is commonly known
as a multilayer perceptron. Although there are several other
architectures and different local operations from the one sketched
above, the multilayer perceptron is by far used the most for
classification tasks.

The most popular way to adjust the free parameters (the strength or
weight of the synapses), in order to teach the neural network to
accomplish the desired task, is the error backpropagation algorithm by
\citet{rumelhart}, which consists of exploring the error hypersurface
by following the reversed local error gradient. By presenting the
network with a series of examples for which known desired outputs are
available (the training set), the local gradient of the total error
with respect to the connection weights can be computed and the weights
correspondingly updated. There are several techniques to achieve
generalization, understood as the ability of a network to imprint in
its weights the abstract rules for classification implicit in the
training examples, disregarding at the same time the particular
details of the examples used.  Again, the most common practice
consists of dividing the available set of examples into three groups:
a training set, a validation set and a test set. Learning proceeds by
minimizing the training set error while at the same time monitoring
the validation error. When the network has captured the general rules
for classification and started to incorporate the particular details
of the training set, the validation error reaches a minimum while the
training error continues decreasing.  It is at this minimum point that
learning is stopped, in order to avoid overtraining, and the error of
the network is estimated using the error set. There are multiple
variations to this very basic scheme, but most of them end up in the
vicinity of a local minimum of the error hypersurface which we expect
to be the global minimum.

Here we deviate from the common practice and use a different formalism,
which we consider more flexible and sound: Bayesian training of neural
networks. In the Bayesian framework, instead of a class assignement
we obtain a predictive probability distribution. Let $\vec{\theta}$
denote the set of parameters needed to fully specify a neural network
architecture (i.e., all the connection weights between neurons in the
network). The network class prediction ${\cal C}_{n+1}$ for a new test
case $\vec{x_{n+1}}$ given a training set $\vec{{\cal S}_{\rm train}}$
is computed as
\begin{equation}
P({\cal C}_{n+1}|\vec{x_{n+1}},\vec{{\cal S}_{\rm train}}) = \int P({\cal
  C}_{n+1}|\vec{x_{n+1}},\vec{\theta}) \cdot P(\vec{\theta}|\vec{{\cal
  S}_{\rm train}}) \cdot d\vec{\theta},
\label{mcmc}
\end{equation}
that is, an average of the predictions $P({\cal C}_{n+1} |
\vec{x_{n+1}}, \vec{\theta})$ made by networks covering the whole
$\vec{\theta}$ parameter space, weighted by the posterior probability
of $\vec{\theta}$ given the training set. The expression $P({\cal
C}_{n+1}|\vec{x_{n+1}},\vec{{\cal S}_{\rm train}})$ is a probability
distribution for all possible classes to which $\vec{x_{n+1}}$ can be
assigned, or, equivalently, for all possible values of ${\cal
C}_{n+1}$. The {\em a posteriori} probability can be computed by
applying Bayes theorem
\begin{equation}
P(\vec{\theta}|\vec{{\cal S}_{train}}) \propto P(\vec{{\cal
  S}_{train}}|\vec{\theta})\cdot P(\vec{\theta}),
\label{bayes}
\end{equation}
that is, as the product of the likelihood function and the prior
probability of the network parameters. Once this probability
distribution is obtained, single value predictions can be obtained by
minimization of loss functions, such as squared error loss (equivalent
to guessing the mean) or absolute error loss (equivalent to guessing
the median) or, as in our case, 0-1 loss functions more suitable for
classification tasks (equivalent to guessing the mode).  The integral
in Eq. (\ref{mcmc}) is defined over all parameter space. In the case
of neural networks, this integral is unmanageable without the aid of
special techniques developed for solving similar problems in the
context of theoretical physics. In this work we make use of hybrid
Monte Carlo techniques implemented in the software package {\bf
Flexible Bayesian Methods} by Neal (\citeyear{Neal}). These are used
to approximate the integral by a sum of terms of the form $P({\cal
C}_{n+1} | \vec{x_{n+1}} , \vec{\theta}^{(n)})$, where all the sets of
weights $\vec{\theta}^{(n)}$ follow the probability distribution
$P(\vec{\theta}|{\cal S}_{train})$.  The likelihood function $P({\cal
S}_{train}|\vec{\theta})$ can be a Gaussian function for regression
networks which incorporate noise in the width of the gaussian, or a
softmax model for classification purposes. A full description of
Markov Chain MonteCarlo (MCMC) techniques is clearly beyond the scope
of this paper. We refer the interested reader to classical expositions
of the method, such as \citet{Neal} or \citet{Bishop1995Neural}. We
mention here that the method achieves equilibrium in the target
statistical distribution only after a certain number of networks
$N_{eq}$ have been generated.  Therefore, in general, only networks
created after $N_{eq}$ are used to estimate the integral.

%Why? ARD 

This approach presents several advantages over traditional maximum
likelihood methods such as error backpropagation. The main advantage
stems from the fact that predictions are not formulated in terms of a
unique network but as an average over all networks. Those networks
with larger {\em a posteriori} probabilities contribute more to the
average than the rest implying that it is no longer necessary to limit
the model complexity.

As mentioned above, in classical backpropagation training, model
complexity is usually limited in order to avoid overtraining because
complex models can incorporate increasingly complex features of the
input space, including random noise in the training set. For each
training set size and statistical distribution of patterns there is an
optimal model complexity that is usually sought by cross-validating
the training performance with an independent set of examples called a
validation set. By stopping the learning algorithm at the minimum of
the validation error curve, we are effectively limiting the average
norm of the weight vectors, thus limiting model complexity (see
e.g. \citeauthor{Bishop1995Neural}, \citeyear{Bishop1995Neural}). This
is at the expense of reducing the available set of training examples
to create the validation set. In the Bayesian framework, on the
contrary, if the model and prior probabilities (henceforth priors) are
appropriate, the inferences are right independent of the training set
size, thus eliminating the need for cross validation and for the
reduction of the training set implied by it.

One of the main advantages of Bayesian training of neural networks is
the possibility of includincluding hierarchical priors in
Eq. (\ref{bayes}).  It allows automatic relevance determination (ARD)
of the parameters by introducing correlations amongst groups of
parameters, in particular, amongst the set of weights connecting a
given input unit with neurons in the next layer. A prior specification
for the network parameters $\vec{\theta}$ can be expressed as the
product of several independent fully specified probabilities (one for
each connection weight at the lowest level) or as the integral of a
more general probability distribution that applies to the connection
weights of a given unit and that is characterized by new sets of
parameters.  Because these newly introduced parameters directly
dictate not the weight probability density but the probability
distribution of the parameters that describe it (i.e., that describe
the weight probability density), they are called hyperparameters. In
the first case, a Gaussian prior with fixed mean and width can be used
directly for the probability distribution of a given connection
weight.  In the second approach, this probability distribution of
weights in the network would be the result of averaging over all
possible means and widths (hyperparameters) weighted by their
respective prior probabilities:
\begin{equation}
P(\vec{\theta}) = \int P(\vec{\theta}|\gamma )\cdot P(\vec{\gamma})
\end{equation}
where, in the example, $\vec{\gamma}$ is the vector of hyperparametric
means and/or widths that is common for all synaptic weights out of the
neuron.  $P(\vec{\gamma})$ in turn can be fully specified or else
given in terms of new hyperparameters at the next level of neural
units in the same layer.  By using different levels of hyperparameters
from the bottom levels of single connection weights or single unit
weight sets up to the highest level of layers or the entire network,
correlations amongst parameter sets of the same group can be
introduced in the integral of Eq.(\ref{mcmc}).

This scheme, when applied to the input-hidden connection weights, can
be used to test the relevance of input variables for the
classification task. If a given input variable is not relevant for
classification purposes, under very special circumstances it may
worsen the network performance.  By making use of these hierarchical
priors, we can effectively remove noninformative input units simply
because a large fraction of the parameter space with significant
contributions to the integral on the right hand side of Eq.
(\ref{mcmc}) will come from networks with their connection weights set
to zero.

\subsection{Preprocessing of Hipparcos training patterns}

We have used light curves from the Hipparcos catalogue as training
set. As usual with neural networks, the raw data (originally in the
JD-V magnitude space) need to be preprocessed to optimize the
performance of the network. The preprocessing of light curves consists
of several distinct stages briefly summarized here to serve as a guide
for the following explanations.
\begin{enumerate}
\item{Unit conversion and binning}
\item{Pattern completion}
\begin{enumerate}
\item{Pattern regression}
\item{Normalization}
\item{SOM consultation}
\item{Second order interpolation}
\end{enumerate}
\item{Pattern regression}
\item{Normalization and phase-shifting}
\end{enumerate}
\subsubsection{Unit conversion and binning}
First, original JD-V magnitude light curves from the
Hipparcos archive are extracted and observations with bad quality
flags removed. Then, the time coordinate is converted into phase
according to Hipparcos ephemeris, if available. Otherwise, literature
ephemerides provided with the catalogue are used. The resulting light
curve is binned into 50 phase intervals between 0 and 1 corresponding
to $\Delta \phi = 0.02$. In a high fraction of the catalogue, light
curves contain gaps in the phase coverage of the binary cycle.
 
The analysis in terms of the information content (IC) of the inputs
supports the choice of 50 bins as a compromise between maximum
possible resolution with manageable sizes. As a rule, very narrow
phase bins can preserve finer details of the light curve.  In practice
however, there is a limit above which fine details convey no useful
information for the classification task. We found that the smallest
detail necessary for the classification task in our classification
scheme was given by the eccentricity definition (see Sect.
\ref{classification}). At a resolution of $\Delta \phi = 0.02$, a
system is classified as eccentric if the secondary eclipse is more
than two bins/input units away from phase $\phi=0.25$.

\begin{figure}[]
\begin{center}
\includegraphics[scale=0.7]{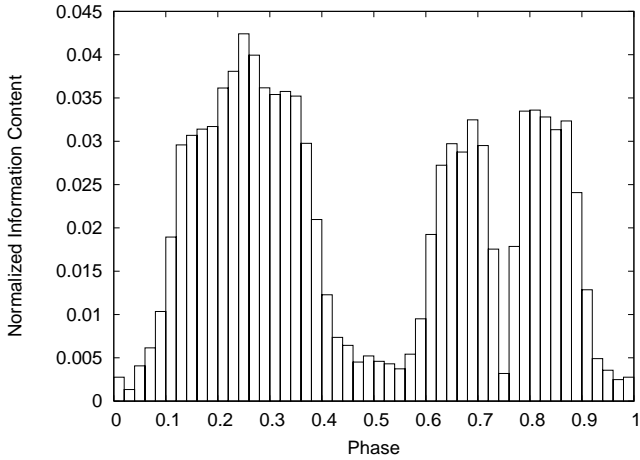}
\end{center}
\caption{\label{mutual-info} Normalized mutual information between
each of the phase bins and the class.}
\end{figure}

Bearing that in mind, we studied the information content distribution
along the light curve. We defined the information content of a given
light curve as the sum of the mutual information content of the
measured phase bins and the class. The mutual information $I$ between
two random variables $X$ and $Y$ was defined as
\begin{equation}
I(X,Y)=\sum_{x,y} p(x,y) \log_2 \frac{p(x,y)}{p(x)\cdot p(y)}.
\end{equation}
We have computed the mutual information between each of the 50 phase
bins ($X_i, 1<i<50$) and the class category ($Y$). The resulting
distribution is shown in Fig. \ref{mutual-info}. Equivalent plots
at higher resolutions (i.e., with smaller $\Delta\phi$) do not change
this smooth curve. 

This plot shows that at a resolution of 50 bins, the mutual
information curve is smooth and intuitively reflects the knowledge
needed by a human classifier. At much higher resolutions (which do not
convey more information), there are not enough examples to
characterize the relationship between each bin mean and the class, and
the performance of the neural networks degrades. At lower resolutions,
important information is lost and, again, the performance degrades,
especially when identifying eccentricity.

\subsubsection{Pattern completion}
Although neural networks are characterized by a high fault tolerance
that can be assimilated to the presence of gaps in the phase coverage,
we observed improved performance of the classification network when
these gaps are interpolated according to the procedure presented
below, mainly in cases where incomplete phase coverage is worsened by
the presence of significant noise and/or outliers.  Therefore, chose
to interpolate data in the gaps by applying the following
approach. The incomplete light curve is presented to a Self Organized
Kohonen Map (SOM) constructed with the best and most representative
721 light curves of the Hipparcos catalogue, including pulsating
variables (see below). These were chosen to cover as many
morphological features as possible with low noise levels and a phase
coverage such that a simple spline interpolation allows reliable
recovery of all morphological information. Following presentation of
the incomplete light curve to the SOM, the map of neural activity is
searched for the most similar light curve of the map that uses the
Euclidean distance criterion.

%SOM creation
The SOM was created using the standard software package {\bf SOM\_PAK}
prepared by the SOM programming team of the Helsinki University of
Technology. The map had dimensions 10 by 8, it was trained 100 times
with different initializations, and the lowest quatization error map
was saved for subsequent use. During training, Gaussian neighbourhood
functions were used. The automatic procedure implemented in {\bf
SOM\_PAK} for the search of such minimum error nets implies a random
initialization and two training stages: during the first 1000 cycles,
unit vectors are ordered in a process whereby the neighbourhood radius
decreases from values close to the map size down to unity, and the
learning rate decreases from 0.05 to zero. During the next 10000
cycles, unit vectors are fine-tuned to minimize quantization error by
training with smaller rates (starting at 0.02) and neighbourhood radii
(starting from 3.0). The choice of the map dimensions is justified in
terms of the Sammon mapping of the input set \citep{Sammon:1969a}.

Presentation of an incomplete light curve to the SOM requires adequate
preprocessing. In this case, the preprocesing consists of normalizing
the light curve considered as a 50-component vector to unit length, as
done with the map creation vectors. The reason for this is that a SOM
operates by computing the scalar product of the input vector and each
of the neural codebook vectors, thus constituting a morphological
similarity detector. Therefore, it is necessary to scale the input's
incomplete light curve vector to a length at least close to the one
used for the codebook vectors. Unfortunately, the normalization
constant of an incomplete light curve will be smaller than if it were
complete, by a factor that depends on the gap total length and the
precise phases missing from the curve. Thus, in order to properly
normalize the incomplete light curve, we need the same phase bins that
we want to retrieve from the SOM. To overcome this difficulty, the
original data previous to the phase-binning process are regressed
using a set of neural networks obtained under the Bayesian framework
described above.

%Details of the first regression

The regression network is indeed a set of networks, the parameters of
which follow the distribution function $P(\vec{\theta}|\vec{{\cal
S}_{train}})$, with $\vec{{\cal S}_{train}}$ the set of points in the
light curve. This set of networks is generated by specifying priors
with hyperparameters for input-to-hidden weights, hidden biases,
hidden-to-output weights, and output biases.

The prior specification used for the output bias is a Gaussian prior
with a mean of zero and standard deviation 10. For input-to-hidden
weights and hidden biases, a Gaussian distribution is used with zero
mean and variance given by a gamma distribution of mean equal to $2.0$
and $\alpha=0.5$, where alpha is the shape parameter. Finally, the
hidden-to-output weights are given Gaussian priors with mean equal to
$3.0$ and $\alpha=0.5$. These weights are automatically rescaled based
on the number of hidden units so that the effect is independent of the
hidden layer size in the limit of large numbers. Again, we refer those
readers interested in the details of this method to the classical
exposition by \citet{Neal}.

This regression network is then used to interpolate the missing gaps
and the result is used as the basis for computing the normalization
constant. It is important to notice that the regression is only used
for normalization purposes. The query to the SOM is made with the
incomplete light curve, previously normalized to unity with the length
derived from the light curve completed by regression.

Once the SOM has been consulted and the resulting neural activity map
searched for the closest exemplars, these are used to fill in the gaps
of the incomplete input. The process modifies the zero and first-order
terms of the retrieved exemplar in the missing phase interval (i.e.,
adds a linear function) to match the limiting data, and it uses only
higher order curvature terms. The correction applied to the values of
the retrieved exemplars in the phase gaps are given by
\begin{equation}
V_i = V^{ex}_i\cdot (\alpha + m\cdot (\phi_{i2}-\phi_{i1}))
\end{equation}
where
\begin{eqnarray}
m &=& (\beta - \alpha) / (\phi_{i2}-\phi_{i1})\\
\alpha &=& V_{i1}/V^{ex}_{i1}\\
\beta &=& V_{i2}/V^{ex}_{i2},
\end{eqnarray}
and $i1$ is the subindex of the last sampled phase bin before the gap,
$i2$ is the subindex of the first sampled phase bin after the gap,
$V_{i1}$ and $V_{i2}$ are the values of the average magnitudes
measured in the corresponding phase bins of the incomplete curve, and
$V^{ex}_{i1}$ and $V^{ex}_{i2}$ the magnitudes in bins $i1$ and $i2$
of the exemplar curve.

\subsubsection{Pattern regression}
The result of the SOM-based pattern completion ($V$ magnitudes for
phase bins not sampled) is added to original data obtained from the
catalogue before binning in phase. The completed light curve is
regressed again using a second set of neural networks totally
equivalent to the one used in the pattern completion stage, and the
result is used to obtain an equispaced light curve of, again, 50 bins.

\subsubsection{Normalization and phase-shifting}
Finally, the result is shifted in phase, to make the minimum of the
light curve (maximum magnitude) coincide with $\phi=0.75$, and rescaled
in magnitude to match the $[0,1]$ interval. This final product is used
both in the training of the classification network and as input to the
trained classifier.

\subsection{\label{classification}Classification}

A total number of 1722 light curves were used for the training of the
network. The relative size of each group of curves is given in Table
\ref{statistics}. In it, we split the set of Type 1 light curves to
create a new group of systems (named Type 0) with essentially the same
detached configuration made explicit in Sect.  \ref{2}, but with
$\Delta \phi$ between maxima greater than 0.29 or less than 0.21,
i.e., systems with clearly detectable eccentricies with the phase bin
width used in the preprocessing. Of the 1722, 1610 were directly taken
from the Hipparcos catalogue, while the remaining 112 are synthetic
light curves of eccentric systems covering all possible $\Delta \phi$
between eclipses in steps of one phase bin, and depths of the
secondary eclipse relative to the primary of 100, 80, 60, and
40\%. This addition to the basic Hipparcos training set was included
to improve the poor performance of the classifier as an eccentricity
detector when trained only with Hipparcos light curves, basically due
to the scarcity of eccentric binary systems. Given that eccentric
binaries only represent a small percentage of the total 1610 light
curves, the overall performance of the network did not improve (it
even degraded from a 6.1 \% average misclassification rate to 6.9 \%),
but the misclassification rate that was restricted to the eccentric
systems lowered from an average 20\% down to 4\%.  The advantage
introduced by the new class of eccentric systems is the possibility of
using this classifier as a first stage in the automatic generation of
lists of pre-main sequence binary system candidates in which the light
curve information is combined with spectral or colour data.

\begin{figure*}[]
%\begin{center}
\includegraphics[scale=0.25,angle=-90]{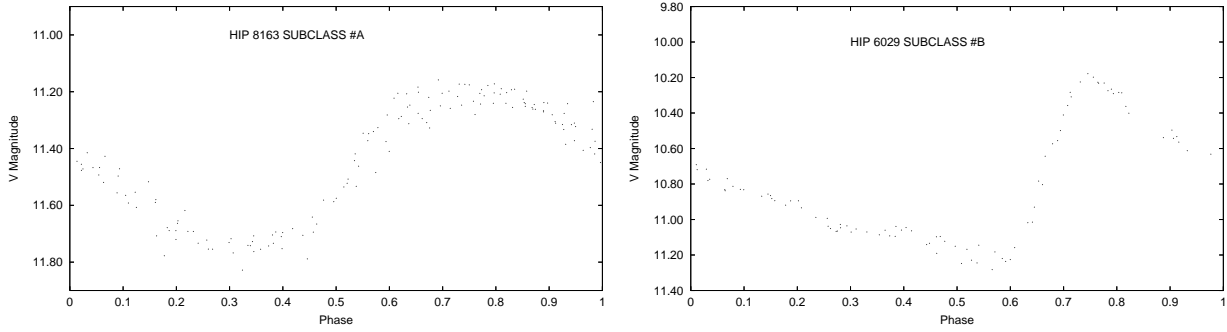}
%\end{center}
\caption{\label{pulsators} Two examples of the new classes used to separate 
  pulsating star light curves from binary systems.}
\end{figure*}

Besides splitting Type 1 systems, the table includes two new groups of
light curves corresponding to pulsating variables light curves of two
morphologies: sine-like curves with symmetric ascending and descending
slopes (type A) and asymmetric light curves (type B).  Figure
\ref{pulsators} shows light curves from the Hipparcos catalogue as
examples of both new types of pulsating morphologies. The reason for
this noninformative classification is that morphological information
alone is not enough to separate the different classes of pulsating
stars. As mentioned above, this is the subject of ongoing research.
The exact statistics of the pulsating stars light curves used for
training are given in Table \ref{stat-puls}.

\begin{table*}[]
\caption[]{Number of light curves of each class used in the training set. Type 0 includes eccentric systems from the Hipparcos catalogue plus 112 synthetic light curves (see text).}
\label{statistics}
\centering
\begin{tabular}{c c c c c c c}
%\hline
Type 0 & Type 1 & Type 2 & Type 3 & Type 4 & Type A & Type B\\
\hline
\hline
32(+112) &269 & 164 & 192 & 129 & 131 &  693 \\
\hline
\end{tabular}
\end{table*}

\begin{table*}[]
\caption[]{Number of light curves of each class of pulsation used in the training set 
(pulsation class taken from the Hipparcos catalogue).}
\label{stat-puls}
\centering
\begin{tabular}{c c c c c c c c c}
%\hline
& $\alpha$ Cygni & $\beta$ Cephei & Cepheids & W Virginis & $\delta$
  Cepheids & $\delta$ Scuti & Mira & RR Lyrae \\
\hline
\hline
Type A & 5 & 26 & 2 & 2 & 20 & 51 & 19 & 6 \\
Type B & 11 & 20 & 17 & 23 & 226 & 36 & 182 & 178 \\

\hline
\end{tabular}
\end{table*}

In this case, the output bias was given a Gaussian distribution of
zero mean and variance given by a hyperparameter taken from a gamma
distribution of mean $0.05$ and shape parameter $0.5$. This was also
the case for the (rescaled) hidden-to-output weights and the hidden
biases. Input-to-hidden weights were given a higher level of
hyperparameters: their values were taken from a Gaussian distribution
of zero mean and a hyperparameterized variance; this variance in turn
was given for all such weights of a given input unit by a gamma
distribution of shape parameter $0.5$ and mean given by the overall
gamma distribution (common for all input units) of mean $0.2$ and
shape parameter $0.5$. This hierarchical scheme introduces
dependencies in the values of the weights connecting a given input
unit to the hidden layer, thus allowing automatic (implicit) pruning of
the input unit, if it does not add a significant improvement to the
performance of the net. Finally as mentioned above, a softmax model is
used in which the probability that the input light curve $\vec{x}$ is
of class $k$ is defined in terms of the output of the network as
\begin{equation}
P(C=C_{k}|\vec{x})=\frac{\exp(f_k(\vec{x}))}{\sum_{k'}f_{k'}(\vec{x})},
\end{equation}
with $f_{k'}$ the output of neuron $k'$ that represents class
$C_{k'}$.

The integral in Eq. (\ref{mcmc}) is approximated by the sum of 200
terms generated by the Markov Chain MonteCarlo method after the
distribution is let stabilize in 800 initial steps. Visual inspection
of the evolution of the error and weights confirms that, even before
600 initial steps, the method attains an equilibrium distribution. The
errors reported in the next section have the same statistical
properties if computed with the last 300 or 400 networks generated.

\onecolumn

{\scriptsize
\begin{center}
\tablecaption{Relevant physical parameters of systems classified as class 1 in the text. }
\label{Tipo1}
\noindent

\tablefirsthead{\multicolumn{11}{c}{} \\
\multicolumn{11}{c}{ }\\ \hline\hline
Name & $P(d)$     &$q$   &Comp.  &Spec. &M           &R
&T$_{\rm eff}$ &log(L) & $M_{V}$    &ref.\\
HD     & V$_{\rm max}$ &a(R$_{\odot})$ & &Type &(M$_{\odot})$
&(R$_{\odot})$      &(K)         &L($_{\odot}$) & &\\
\hline\hline
}

\tablehead{\multicolumn{11}{l}{\footnotesize \it ......continued from
    previous page.}\\ \hline\hline 
Name & $P(d)$     &$q$   &Comp.  &Spec. &M           &R
    &T$_{\rm eff}$ &log(L)     & $M_{V}$ &ref.\\
HD     & V$_{\rm max}$ &a(R$_{\odot})$ & &Type &(M$_{\odot})$
&(R$_{\odot})$      &(K)         &L($_{\odot}$) & &\\

\hline
}

\tabletail{\hline 
\multicolumn{11}{r}{\footnotesize \it Continued on the next page...}\\
}
\tablelasttail{\multicolumn{11}{c}{}
}

\begin{supertabular}{|lllllllllll|}

BW Aqr    &6.72          &0.931   &A   &F7V       &1.488  &2.064     &3.803   &0.79    &2.74 &1,2 \\
BD $-16^{\circ}6074$  &10.33 & 21.298   &B   &F5V       &1.386  &1.788     &3.810   &0.70    &2.98 &\\
\hline  
V539 Ara  &3.17    &0.851  &A   &B3V       &6.254  &4.432     &4.260   &3.29   &-1.70 &1,3,4,5\\ 
HD 161783  &5.71   &20.53  &B   &B4V       &5.326  &3.734     &4.230   &3.02   &-1.04 &\\
\hline
EM Car    &3.41   &0.936   &A   &08V        &22.89  &9.34      &4.531   &5.02    &4.56 &1,6\\
HD 97484   &8.38  & 33.72     &B   &08V     &21.43  &8.33     &4.531   &4.92    &4.31 &\\
\hline
GL Car    &2.42   &0.962   &A   &B8V      &13.5   &4.99      &4.476   &5.02   &-2.97 &7\\
HD 306168  &8.38  &22.60   &B   &B8V       &13.0   &4.74       &4.468   &4.92   &-2.83 &\\
\hline
QX Car    &4.48   & 0.915   &A   &B2V      &9.267  &4.289    &4.377   &3.72   &-2.32 &1,4,8\\ 
HD 86118   &6.64  & 29.81   &B   &B2V       &8.480  &4.051     &4.354   &3.58   &-2.07 &\\ 
\hline
SZ Cen    &4.11   &0.982    &A   &A7V     &2.317  &4.554     &3.875   &1.77    &0.29 &1,9,10\\ 
HD 120359  &8.48 &17.94    &B   &A7V     &2.277  &3.624     &3.892   &1.64    &0.61 &\\
\hline
CW Cep    &2.73   &0.893  &A   &BO.5V     &13.52  &5.685     &4.452   &4.27   &-3.17 &1,11,12\\
HD 218066  &7.59  &24.217   &B   &BO.5V      &12.08  &5.177     &4.442   &4.15   &-2.94 &\\
\hline
EK Cep    &4.43   &0.553  &A   &Al.5V     &2.029  &1.579     &3.954   &1.17    &1.89 &1,11,13,14,15\\
HD 206821  &7.87  &16.63   &B   &G5Vp      &1.124  &1.315     &3.756   &0.21    &4.31 &\\
\hline
RZ Cha    &2.83   &0.994  &A   &F5V       &1.518  &2.264    &3.810   &0.90    &2.46 &1,10,16\\
HD 93486   &8.10   &12.17  &B   &F5V       &1.509  &2.264     &3.810   &0.90    &2.46 &\\
\hline
V442 Cyg  &2.39   &0.901  &A   &FIV       &1.564  &2.072    &3.839   &0.94    &2.35 &1,17\\
HD 334426  &9.72  &10.81   &B   &F2V       &1.410  &1.662    &3.833   &0.72    &2.89 & \\
\hline
V1143 Cyg &7.64   &0.968  &A   &F5V       &1.391  &1.346     &3.810   &0.45    &3.60 &1,10,11 \\
HD 185912  &5.86  &22.82   &B   &F5V       &1.347  &1.323     &3.806   &0.42    &3.67 & \\
\hline
DI Her    &10.55  &0.874 &A   &B5V        &5.185  &2.680     &4.230   &2.73    &-0.46&1,18 \\
HD 175227  &8.42  &43.18   &B   &B5V       &4.534  &2.477     &4.179   &2.46    &4.05 & \\
\hline
RX Her    &1.78   &0.847  &A   &B9        &2.75   &2.44      &4.015   &1.79    &0.48 &10,11,19\\
HD 170757  &7.26  &10.62  &B   &A0         &2.33   &1.96     &3.985   &1.48    &1.12 &\\
\hline
Al Hya    &8.29   &0.922 &A   &F2m        &2.145  &3.914     &3.826   &1.44    &1.10 &1,20 \\
BD $+0^{\circ}2259$ &9.36  &27.630   &B   &FOV      &1.978    &3.850   &3.851   &1.24    &1.61 & \\
\hline
TZ Men    &8.57   &0.604 &A   &A0V        &2.487  &2.016   &4.017   &1.63    &0.93 &1,21\\
HD 39780   &6.19  &27.94  &B   &A8V        &1.504  &1.432    &3.857   &0.69    &2.97 & \\
\hline
UX Men    &4.18   &0.967    &A   &F5V        &1.238  &1.347     &3.792   &0.38    &3.81 &10,22,23 \\
HD 37513   &8.22  &14.68  &B   &F8V       &1.198  &1.274     &3.789   &0.32    &3.96 & \\
\hline
V451 Oph  &2.20    &0.848 &A   &B9V       &2.776  &2.640     &4.033   &1.93    &0.34 &1,10,11,24\\
HD 170470  &7.87   &12.27 &B   &A0V       &2.356  &2.028    &3.991   &1.53    &1.11 &\\
\hline
V1031 Ori &3.41    &0.936  &A   &A6V        &2.473  &4.321     &3.895   &1.80    &0.18 &1,25\\
HD 38735   &6.02   &33.727 &B   &A3V        &2.286  &2.977     &3.924   &1.60    &0.74 &\\
\hline
Al Phe    &24.59   &0.966 &A   &KOIV     &1.236  &2.930     &3.700   &0.69    &3.24 &1,26\\
HD 6980    &8.61   &47.830  &B   &F7V       &1.195  &1.816     &3.800   &0.67    &3.07 & \\
\hline
Zeta Phe  &1.67    &0.649  &A   &B6V      &3.930  &2.851     &4.163   &2.51    &4.59 &1,3,10,27\\
HD 6882    &3.95   &11.039  &B   &B8V       &2.551  &1.853     &4.076   &1.79    &0.91 &\\
\hline
V1647 Sgr &3.28    &0.900  &A   &AIV      &2.189  &1.831     &3.982   &1.41    &1.35 &1,28\\
HD 163708  &6.94   &14.93   &B   &AIV      &1.972  &1.666     &3.959   &1.23    &1.73 &\\
\hline
V760 Sco  &1.73    &0.927 &A   &B4V       &4.980  &3.013     &4.228   &2.82    &4.71 &1,29\\
HD 147683  &6.99   &12.88  &B   &B4V       &4.620  &2.640     &4.210   &2.63    &4.24 &\\
\hline
CV Vel    &6.89    &0.982  &A   &B2.5V   &6.100  &4.087     &4.253   &3.19    &-1.48 &1,30\\
HD 77464   &6.69   &34.96  &B   &B2.5V     &5.996  &3.948     &4.250   &3.15    &-1.38 &\\
\hline\hline
\end{supertabular}
\end{center}
}
\vspace{1cm}
{\footnotesize {\bf References used in Table \ref{Tipo1}:} 
1:~\citet{And1991}, 
2:~ \citet{Clau1991},
3:~ \citet{Ande1983}, 
4:~ \citet{DeGr1989}, 
5:~ \citet{Clau1996}, 
6:~ \citet{Ande1989},
7:~ \citet{Gime1986}, 
8:~ \citet{AndClaNor1983}, 
9:~ \citet{Gron1977},   
10:~ \citet{Popp1980}, 
11:~ \citet{Frac1972}, 
12:~ \citet{ClaGim1991}, 
13:~ \citet{Pop1987},
14:~ \citet{MarReb1993}, 
15:~ \citet{ClaGimMar1995}, 
16:~ \citet{Jorg1975},
17:~ \citet{Lacy1987}, 
18:~ \citet{Popp1982}, 
19:~ \citet{Jeff1980},
20:~ \citet{Khal1989}, 
21:~ \citet{AndClaNor1987}, 
22:~ \citet{ClaGro1976},
23:~ \citet{AndClaMag1989}, 
24:~ \citet{ClaGimSca1986},
25:~ \citet{AndClaNor1990}, 
26:~ \citet{AndClaNor1988},  
27:~ \citet{ClaGylGro1976}, 
28:~ \citet{ClaGylGro1977},
29:~ \citet{AndClaNor1985}, 
30:~ \citet{ClaGro1977}
} 

\newpage
{\scriptsize
\begin{center}
\tablecaption{Relevant physical parameters of systems classified as class 2 in the text.}\label{Tipo2}
\tablefirsthead{\multicolumn{11}{c}{} \\
\multicolumn{11}{c}{ }\\ \hline\hline
Name & $P(d)$     &$q$   &Comp.  &Spec. &M           &R
&T$_{\rm eff}$ &log(L)     & $M_{V}$ &ref.\\
HD     & V$_{\rm max}$ &a(R$_{\odot})$ & &Type &(M$_{\odot})$
&(R$_{\odot})$      &(K)         &L($_{\odot}$) & &\\
 \hline\hline
}
\tablehead{\multicolumn{11}{l}{\footnotesize \it ...continued from
    previous page.}\\ 
\multicolumn{11}{c}{ }\\ \hline\hline
Name & $P(d)$     &$q$   &Comp.  &Spec. &M           &R
&T$_{\rm eff}$ &log(L)     & $M_{V}$ &ref.\\
HD     & V$_{\rm max}$ &a(R$_{\odot})$ & &Type &(M$_{\odot})$
&(R$_{\odot})$      &(K)         &L($_{\odot}$) & &\\

\hline
}

\tabletail{\hline\multicolumn{11}{r}{\footnotesize \it Continued on
    the next page...}\\ }

\tablelasttail{\hline}

\begin{supertabular}{|lllllllllll|}

RY Aqr    &1.9666  &0.204   &A   &A3         &1.27   &1.28      &3.881   &0.700   &2.9  &1,2\\
HD 203069    &9.06  & 7.61     &B   &           &0.26   &1.79      &3.658
&0.100   &4.4  &\\\hline

IM Aur    &1.2473  &0.249   &A   &B9         &4.73   &3.24      &4.199   &2.770   &-2.2  &1,3,4\\
HD 33853     &7.70 &8.81   &B   &          &1.18   &2.20      &3.881   &1.160   &1.8  &\\\hline

R  CMa    &1.1359  &0.131  &A   &F2V         &1.52   &1.73       &3.849   &0.76    &2.77  &3,5\\
HD 57167     &4.5730 &5.48    &B   &G8IV       &0.20   &1.18        &3.712   &-0.41   &6.36  &\\\hline

RZ Cas    &1.195   &0.330  &A   &A3V         &2.21   &1.67     &3.934   &1.12    &  &3,6,7\\
HD 17138     &6.2  &6.79      &B   &           &0.73   &1.94      &3.672   &0.16    &  &\\\hline

TV Cas    &1.8126  &0.464   &A   &B9V        &2.80   &2.81     &4.029   &1.970   &-0.2  &1,3,7,8\\
HD 1486      &10.57   &10.01   &B   &G5-9IV     &1.30   &3.15    &3.708   &0.780   &2.7  &\\\hline
   
U CrB     &3.4522  &0.300  &A   &B6V         &4.70   &2.60     &4.185   &2.520   &-1.6  &1,3,7,9,10\\
HD 136175    &7.65  &17.57    &B   &GOIII-IV    &1.41   &4.91     &3.764   &1.390   &1.2  &\\\hline
          
SW Cyg    &4.5730  &0.220  &A   &A2V         &2.27   &2.43    &3.957   &1.550   &0.8  &11\\
HD 191240    &9.30 &16.28      &B   &KOIV       &0.50   &4.15     &3.690   &0.950   &2.3  &\\\hline
           
AF Gem    &1.2435  &0.342  &A   &B9.5V       &3.37   &2.61      &4.00    &1.78    &  &7,12\\
HD 210892    &10.54   &8.04  &B   &G0III-IV    &1.155  &2.32      &3.768   &0.75    &  &\\\hline

AQ Peg    &5.5485  &0.256   &A   &A2         &2.34   &2.64     &3.959   &1.630   &0.6  &1,11\\
BD $+12^{\circ}4653$  &10.39   &18.887  &B   &       &0.60   &4.89    &3.644   &0.910   &2.4  &\\\hline

AT Peg    &1.1461  &0.472   &A   &A4V        &2.22   &1.86      &3.924   &1.19    &1.76  &1,3,7,13\\
HD 210892    &9.50   &6.84         &B   &          &1.05   &2.15      &3.690   &0.38    &4.1  &\\\hline

AW Peg    &10.6225 &0.160  &A   &AlVe        &2.06   &1.90     &3.959   &1.350   &1.3  &1,3,11,14\\
HD 207956    &7.40   &27.18   &B   &F5IV        &0.33   &6.12     &3.602   &0.930   &2.4  &\\\hline
 
DM Per    &2.7277  &0.314   &A   &B5V       &5.82   &3.96     &4.202   &2.960   &-2.6  &1,3,15,16\\
HD 14871     &7.88   &16.18    &B   &A5III      &1.83   &4.60     &3.914   &1.940   &-0.1  &\\\hline

RY Per    &6.8636  &0.281  &A   &B3V         &6.60   &4.00     &4.246   &3.140   &-3.2  &1,17,18\\
HD 17034     &8.48    &30.96   &B   &F6IV      &1.86   &8.53     &3.814   &2.070   &-0.5  &\\\hline
          
Bet Per   &2.8673   &0.221   &A   &B8V       &3.70   &2.90     &4.097   &2.250   &-0.9  &1,3,19,20\\
HD 19356     &2.12    &14.04  &B   &G8-KOIII    &0.82   &3.50     &3.708   &0.860   &2.5  &\\\hline
          
U Sge     &3.3806  &0.333   &A   &B8V        &5.70   &4.05     &4.146   &2.750   &-2.2  &1,9,18,21\\
HD 181182    &8.20    &18.63   &B   &G4III      &1.90   &5.38     &3.724   &1.310   &1.4  &\\\hline
           
XZ Sgr    &3.2756  &0.137   &A   &A3V        &1.89   &1.46     &3.945   &1.060   &2.0  &1,22\\
HD 168710    &0.00  & 11.98     &B   &G5IV      &0.26   &2.47     &3.708   &0.570   &3.3  &\\\hline

TX UMa    &3.2756  &0.137   &A   &B8V        &4.76   &2.83      &4.111   &2.30   &  &1,3,9,23\\
HD 93033     &7.06  &11.98    &B   &G0III-IV    &1.18   &4.24      &3.740   &1.17   &  &\\\hline\hline
\end{supertabular}
\end{center}
}
\vspace{1cm}
{\footnotesize {\bf References used in Table \ref{Tipo2}:} 
1:~ \citet{GiuMarMez1983}, 
2:~ \citet{Helt1987},
3:~ \citet{Frac1972}, 
4:~ \citet{GulSez1985}, 
5:~ \citet{SarVivAbh1996}, 
6:~ \citet{MaxHilHil1994},
7:~ \citet{MaxHil1996}, 
8:~ \citet{KhaHil1992}, 
9:~ \citet{CesFedGiu1977},   
10:~ \citet{Gent1989}, 
11:~ \citet{WilMuk1988}, 
12:~ \citet{MaxHil1995}, 
13:~ \citet{MaxHilHil1994},
%15:\citet{GulGudSez1993}, 
14:~ \citet{DerDem1992}, 
15:~ \citet{HilSkiCar1986},
16:~ \citet{HilHil1992}, 
17:~ \citet{OlsPla1997},
18:~ \citet{VHamWil1986},
19:~ \citet{RicMocBol1988},
20:~ \citet{Kim1989}, 
%21:\citet{AlbGeaRic1995}, 
21:~ \citet{KhaBud1986},
22:~ \citet{Kni1974},
23:~ \citet{MaxHilHil1995}, 
} 

\newpage
{\scriptsize
\begin{center}

\tablecaption{Relevant physical parameters of systems classified as class 3 in the text.}\label{Tipo3}
\tablefirsthead{\multicolumn{11}{c}{} \\
\multicolumn{11}{c}{ }\\ \hline\hline
Name & $P(d)$     &$q$   &Comp.  &Spec. &M           &R
&T$_{\rm eff}$ &log(L)     & $M_{V}$ &ref.\\
HD     & V$_{\rm max}$ &a(R$_{\odot})$ & &Type &(M$_{\odot})$
&(R$_{\odot})$      &(K)         &L($_{\odot}$) & &\\
\hline\hline
}

\tablehead{\multicolumn{11}{l}{\footnotesize \it ...continued from
    previous page.}\\ 
\multicolumn{11}{c}{}\\ \hline\hline 
Name & $P(d)$     &$q$   &Comp.  &Spec. &M           &R
    &T$_{\rm eff}$ &log(L)     & $M_{V}$ &ref.\\
HD     & V$_{\rm max}$ &a(R$_{\odot})$ & &Type &(M$_{\odot})$
&(R$_{\odot})$      &(K)         &L($_{\odot}$) & &\\

\hline
}

\tabletail{\hline 
\multicolumn{11}{r}{\footnotesize \it Continued on the next page...}\\
}

\tablelasttail{\hline}

\begin{supertabular}{|lllllllllll|}

CX Aqr    &0.5559  &0.537   &A   &F5              &1.19   &1.29       &3.806  &2.70 &&1,2    \\  
          &10.70   &3.48  &B   &G9               &0.64   &1.15        &3.696  &0.72 &    &   \\\hline 
EE Aqr    &0.5089  &0.322    &A   &F0-F2          &2.20   &1.75        &3.881  &0.88 &&1,3,4       \\
HD 213683 &8.30    &3.83  &B   &                 &0.71   &1.07        &3.643  &-0.42 &  &   \\\hline
IU Aur    &1.8115  &0.676   &A   &BOV         &21.3    &7.5      &4.505 &4.73  &-6.8  &5,6,7\\
HD 35652     &8.19    &20.58  &B   &BO.5V        &14.4    &7.2      &4.449 &4.46  &-6.3  &\\\hline    
TT Aur    &1.3327  &0.648   &A   &B2V         &8.58    &4.06     &4.373 &3.664 &-4.5  &8,9\\
HD 33088     &8.53    &12.32   &B   &            &5.56    &4.17     &4.267 &3.264 &-3.5  &\\\hline
DO Cas    &0.6847  &0.313   &A   &A               &1.69   &2.10        &3.96   &1.42 &&1,5,10,11 \\
          &8.60    &4.26    &B   &               &0.53   &1.20        &3.68   &-0.16 &&  \\\hline
YY Cet    &0.79    &0.510   &A   &A8              &1.84   &2.09       &3.875  &1.10 && 1,12  \\   
BD $-18^{\circ}349$ &5.05 &10.00   &B   &   &0.94   &1.63        &3.725  &0.30 &&   \\\hline    
AI Cru    &1.4177  &0.611  &A   &B2IVe        &10.30   &4.95     &4.384 &3.880 &-4.9  &13\\
-60.3723  &9.20    & 13.54 &B   &B4           &6.30    &4.43    &4.248 &3.240  &-3.3  &\\\hline 
V836 Cyg  &0.6534  &0.333   &A   &A3              &2.4    &1.96        &4.00   &1.04 && 1,10,14  \\
HD 203470 &8.59    &4.67  &B   &G                &0.80   &1.24        &3.76   &4.32 &&   \\\hline
RZ  Dra   &0.5508  &0.442    &A   &A5             &1.40   &1.62        &3.911  &  & &1,15 \\
          &10.00   &3.57   &B   &K2              &0.62   &1.12        &3.690  &&  & \\\hline
RU  Eri   &0.6322  &0.420  &A   &F3V             &2.45   &2.06        &       &1.07  &&1,10   \\
HD 24658  &9.90    &4.69   &B   &               &1.03    &1.43        &       &-0.03 &&   \\\hline
u Her     &2.0510  &0.38  &A   &B2IV         &7.60    &5.80     &4.301 &3.680 &-4.5  &5,6,16,17\\
HD 156633    &4.77    &14.87  &B   &B8III        &2.90    &4.40   &4.065 &2.490 &-1.5  &\\\hline
TT  Her   &0.9121  &0.435    &A   &F2V           &1.56    &2.30        &3.960   &1.13&& 1,10,18,19 \\
BD $+17^{\circ}3117$  &5.17  &9.70    &B   &                &0.68    &1.49       &3.744   &-0.02&&   \\\hline
RS Ind    &0.6240  &0.310  &A   &F1V             &2.00    &2.00        &3.857   & 0.98& &1,3,20 \\ 
          &9.90    &4.23  &B   &G8              &0.62    &1.18        &3.668   &-0.23& &  \\\hline
FT Lup    &0.470   &0.426   &A   &F2V            &1.43    &1.43        &3.826   && & 1,21,22\\
132316    &9.7     &3.23   &B   &K5-7V          &0.61    &0.94     &3.639   & & &  \\\hline   
V   Pup   &1.4550   & 0.522  &A   &B1         &14.86   &6.18     &4.450 &4.340 &-6.1  &5,6,23,24\\
HD 65818     &4.41    &15.28  &B   &B3           &7.76    &4.90     &4.425 &4.040 &-5.3  &\\\hline
CX  Vir   &0.7461  &0.336  &A   &F5              &1.07    &1.85        &        &0.75 && 1,25  \\  
123660    &9.20    &3.90  &B   &K               &0.36    &1.12        &        &-0.31& &  \\

\end{supertabular}
\end{center}
}
\vspace{1cm}
{\footnotesize {\bf References used in Table \ref{Tipo3}:} 
1:~ \citet{HilKinMcF1988}, 
2:~ \citet{McFBelAda1986},
3:~ \citet{HilKin1988}, 
4:~ \citet{CovBarMil1990}, 
5:~ \citet{Frac1972}, 
6:~ \citet{GiuMarMez1983},
7:~ \citet{DreHaaLor1994}, 
8:~ \citet{WacPopCla1986}, 
9:~ \citet{Bel1987}, 
10:~ \citet{KarDue1985},
11:~ \citet{BarDiFMil1992}, 
12:~ \citet{McFHilKin1986}, 
13:~ \citet{BelKilMal1987},
14:~ \citet{BreKalKae1989},   
15:~ \citet{KrePajTre1994}, 
16:~ \citet{CesFedGiu1977}
17:~ \citet{Hild1984}, 
18:~ \citet{KweVan1983}, 
19:~ \citet{MilBarMan1989},
20:~ \citet{MarGriLap1990}, 
21:~ \citet{HilKinHil1984}, 
22:~ \citet{LipSis1986},
23:~ \citet{SchDarLeu1979}, 
24:~ \citet{AndClaGim1983},   
25:~ \citet{HilKin1988}, 
}

\newpage
{\scriptsize
\begin{center}

\tablecaption{
 Relevant physical parameters of systems classified as class 4 in the text.}.
\label{Tipo4}

\tablefirsthead{\multicolumn{10}{c}{}\\
\multicolumn{10}{c}{ } \\ \hline\hline
Name & $P(d)$     &$q$   &Comp.  &Spec. &M           &R           &T$_{\rm eff}$ &log(L)     &ref.\\
HD     & V$_{\rm max}$ &a(R$_{\odot})$ & &Type &(M$_{\odot})$
&(R$_{\odot})$      &(K)         &L($_{\odot}$) & \\ 

 \hline\hline
}

\tablehead{\multicolumn{10}{l}
{\footnotesize \it ...continued from previous page.}\\ \hline\hline
Name & $P(d)$     &$q$   &Comp.  &Spec. &M           &R           &T$_{\rm eff}$ &log(L)     &ref.\\
HD     & V$_{\rm max}$ &a(R$_{\odot})$ & &Type &(M$_{\odot})$ &(R$_{\odot})$      &(K)         &L($_{\odot}$) & \\ \hline
}

\tabletail{\hline 
\multicolumn{10}{r}{\footnotesize \it Continued on the next page...}\\
}
\tablelasttail{\hline}

\begin{supertabular}{|llllllllll|}
     
OO   Aql  &0.507  &0.888   &A   &G5V         &1.19   &1.44      &5700    &1.97   &1,2,3,4\\    
HD 187183 &9.20    &4.570  &B   &          &1.34   &1.00        &5635    &1.62   &\\ \hline

V535 Ara  &0.629  &0.582    &A   &A8V       &2.18   &2.10      &8750    &3.17   &1,5\\    
HD 159441 &7.40    &4.667       &B   &         &1.27   &0.79      &8572    &7.86   &\\ \hline

AO   Cam  &0.329  &0.766  &A   &          &1.03   &0.98      &5520    &0.80   &3,6\\    
BD $+52^{\circ}826$     &9.50   &2.452          &B          &0.88   &0.79   &   &5826    &0.80   &\\ \hline

V523 Cas  &0.233  &0.569   &A   &K4         &0.79   &0.75      &4207    &0.16   &3,7,8\\
          &        & 1.714 &B   &          &0.58   &0.45     &4407    &0.12   &\\ \hline

V677 Cen  &0.325  & 0.481  &A   &         &1.06   &1.19      &5745    &1.39   &3,6\\    
          &11.55   & 2.312  &B   &         &0.51   &0.15     &5841    &0.27   &\\ \hline
        
V752 Cen  &0.370  & 0.575  &A   &F8V      &1.20   &1.24      &6210    &2.06   &3,6\\    
HD 101799 &        & 2.681   &B   &        &0.69   &0.36   
   &6234    &0.65   &\\ \hline

VY   Cet  &0.341  &0.666   &A   &         &1.02   &1.01      &5393    &0.77   &9\\    
BD $-20^{\circ}345$ &11.10  &2.449   &B   &         &0.83   &0.68      &5610    &0.61   &\\ \hline

CC   Com  &0.221  &0.518   &A   &         &0.79   &0.41     &4302    &0.17   &2,3,10\\ 
          &11.00   &1.634    &B   &        &0.54   &0.73      &4500    &0.11   &\\ \hline

EK   Com  &0.267  &0.580   &A   &         &0.93   &0.92      &5000    &0.47   &11\\    
          &12.70   &1.981   &B   &         &0.54   &0.28      &5310    &0.20   &\\ \hline

FS   Cra  &0.264  &0.755  &A   &          &0.86   &0.82      &4567    &0.26   &3,10\\    
          &13.80   &1.984  &B   &          &0.73   &0.65      &4700    &0.23   &\\ \hline

YY   Eri  &0.322  &0.693  &A   &G5        &1.01   &1.02      &5389    &0.79   &1,2,3,13\\
HD 26609  &8.80    &2.361  &B   &          &0.70   &0.44      &5585    &0.43   &\\ \hline
SY   Hor  &0.312  &0.659   &A   &         &0.97   &0.95      &4934    &0.47   &3,9\\    
          &11.40   &2.266   &B   &          &0.83   &0.64      &5240    &0.47   &\\ \hline

V508 Oph  &0.345   &0.527       &A   &G5          &1.01   &1.06      &        &0.087   &14\\
BD $+13^{\circ}3496$ &10.00 &2.444  &B   &      &0.52   &0.80      &        &-0.286  &\\ \hline

BB   Peg  &0.362  &0.405  &A   &F8        &1.16   &1.21      &5883    &1.58   &3,15 \\   
          &10.80   & 2.512  &B   &         &0.78   &0.47      &6200    &0.81   &\\ \hline

U   Peg   &0.375   &0.579  &A   &G2V        &1.33   &1.28      &5515    &2.80   &2,3,16 \\   
BD $+15^{\circ}4915$ &9.70  &2.800    &B   &        &0.77   &0.44      &5800    &1.28   &\\ \hline

AE   Phe  &0.362  &0.401  &A    &G1/G2V     &1.17   &1.19      &6000    &1.63   &3,13\\    
HD 9528   &8.30    &2.521    &B    &       &0.79   &0.47      &6145    &0.79   &\\ \hline

YZ   Phe  &0.234  &0.597  &A    &         &0.87   &0.79      &4800    &0.30   &17\\    
          &12.50   &1.786   &B    &        &0.52   &0.35      &5055    &0.16   &\\ \hline

FG   Sct  &0.271  &0.781  &A    &         &0.87   &0.73      &4662    &0.29   &3,10 \\   
          &13.70   &2.036  &B    &         &0.68   &0.83      &4800    &0.25   &\\ \hline

RZ   Tau  &0.416  &0.369  &A    &A7V      &1.57   &1.51      &7200    &5.51   &1,2,18,19\\
HD 285892 &10.50   &3.024  &B    &         &1.00   &0.58      &7146    &2.34   &\\ \hline

BP   Vel  &0.265  &0.722  &A    &         &0.90   &0.86      &4717    &0.33   &20\\   
          &12.90   &2.009 &B    &          &0.65   &0.48      &5000    &0.23   &\\ \hline

BI   Vul  &0.252  &0.686&A    &           &0.86   &0.82      &4549    &0.26   &3,10\\   
          &       &1.898 &B    &           &0.70   &0.59      &4600    &0.20   &\\ \hline

W   UMa   &0.334  & 0.731 &A    &F8V:p    &1.08   &1.10     &5800    &0.87   &1,2,3\\    
HD 83950  &8.30    & 2.505 &B    &         &0.79   &0.51      &6194    &0.60   &\\ \hline

AA   UMa  &0.468  &0.547 &A    &G0        &1.26   &1.40      &5932    &2.17   &3,6\\   
          &11.30   &3.168  &B    &         &1.10   &0.69      &6030    &1.43   &\\ \hline
  
AW   UMa  &0.439 & 0.349  &A    &        &1.52   &1.60      &7175    &6.06   &1,2,18\\   
HD 99946  &7.27    & 3.221  &B    &        &0.53   &0.11      &6875    &0.56   &\\ \hline

RZ   Pyx  &0.656  &0.821  &A    &B7V      &5.76   &2.69      &4.230   &2.73   &21 \\
HD 75920  &8.85    &6.954  &B    &        &4.73   &2.51      &4.225   &2.65   & \\ \hline\hline

\end{supertabular}
\end{center}
}
\vspace{1cm}
{\footnotesize {\bf References used in Table \ref{Tipo4}:} 
1:~ \citet{Frac1972}, 
2:~ \citet{Moc1981},
3:~ \citet{MacVan1996}, 
4:~ \citet{Hriv1989}, 
5:~ \citet{LeuSch1978}, 
6:~ \citet{BarDiFMil1993},
7:~ \citet{Mace1986}, 
8:~ \citet{Same1987}, 
9:~ \citet{LapCla1986},   
10:~ \citet{Bra1985}, 
11:~ \citet{SamGra1995}, 
12:~ \citet{MacVilVan1994}, 
13:~ \citet{LapGom1990},
14:~ \citet{CerMilSca1981}, 
15:~ \citet{ZhaZha1988}, 
16:~ \citet{SamTer1995},
17:~ \citet{WilDev1973}, 
18:~ \citet{MorNaf1997}, 
19:~ \citet{LapGomFar1996},
20:~ \citet{BelMal1987}, 
} 
%%% Local Variables: 
%%% mode: latex
%%% TeX-master: t
%%% End: 

\end{document}